# Generic behavior of ultrastability and anisotropic molecular packing in co-deposited organic semiconductor glass mixtures

*Shinian Cheng[1], Yejung Lee[1], Junguang Yu[2], Lian Yu[2], and Mark D. Ediger[1]*

[1]Department of Chemistry, University of Wisconsin-Madison, Madison, Wisconsin, 53706, USA

[2]School of Pharmacy, University of Wisconsin-Madison, Madison, Wisconsin, 53705, USA

**Corresponding Author**

*E-mail: edigerchem.wisc.edu

**ABSTRACT:** Vapor-deposited glass mixtures of organic semiconductors commonly serve as active layers in organic electronic devices, whose lifetime and performance are strongly influenced by the stability and structure of these mixed glasses. Here, we study the stability and anisotropic molecular packing of six co-deposited organic semiconductor glass mixtures with 50:50 weight ratio, by differential scanning calorimetry and spectroscopic ellipsometry. We find that all six binary systems exhibit high kinetic stability and significantly reduced enthalpy relative to the corresponding liquid-cooled glassy mixtures (ultrastable behavior), even for systems where the glass transition temperatures of the components differ by more than 90 K. Furthermore, we demonstrate that the birefringence of a co-deposited glass mixture, a measure of its anisotropic packing, can be predicted from the birefringence of glasses of the two pure components. These results for stability and structure are expected to be applicable to other co-deposited organic semiconductor glass mixtures, so long as the two components mix well in the glass and individually can form ultrastable glasses. Therefore, our findings are significant for designing novel electronic devices with enhanced device lifetime and increased operational efficiency.

## ▪ INTRODUCTION

Glasses are non-crystalline materials that are widely used in applications where macroscopic homogeneity and smooth surfaces are required. For example, organic semiconductor glasses are utilized as active layers to ensure uniform performance in organic light emitting diode (OLED) displays that are being used in cellphones and televisions[1]. However, as nonequilibrium materials, glasses can evolve with time through physical aging[2], crystallization[3], and degradation[4], which can lead to a loss of device performance[4]. Recent studies have shown that glasses prepared by physical vapor deposition (PVD) can exhibit exceptional kinetic and thermodynamic stability (ultrastable behavior)[5, 6] that can overcome many of these challenges[7-9], broadening their potential use in applications. In addition, PVD glasses can exhibit anisotropic packing[10-12], and the molecular orientation can be continuously tuned from "standing up" to "lying down" relative to the substrate through varying deposition conditions[13]. While most past work has focused on single-component PVD organic glasses[8, 14], devices typically utilize multicomponent PVD glasses, and our understanding of multicomponent systems is much more limited.

One major challenge is to understand the conditions under which co-deposited glasses can exhibit the very high kinetic stability of single-component PVD glasses[5]. Previous studies on single-component PVD glasses have demonstrated that high surface mobility below the glass transition temperature ($T_g$) is the key to forming ultrastable glasses[5]. The surface equilibration mechanism explains that mobility near the surface allows newly deposited molecules to find low energy and highly stable packing arrangements which are then locked into place by subsequent deposition. From this perspective, highly stable co-deposited glass mixtures might be expected at deposition conditions where both components can individually form ultrastable single-component PVD glasses. Recent reports on PVD glass mixtures of isomers[15] and a pair of organic semiconductors with similar $T_g$ values[16] support this viewpoint. However, it is unclear whether this result holds generically for organic semiconductors. For example, would co-deposition of molecules with a large difference in $T_g$ values form ultrastable PVD glass mixtures, since it may be impossible to find a deposition temperature where both components have high surface mobility?

A second important challenge is to understand and control molecular orientation in multicomponent PVD organic glasses. As one example where this is important, the light-emitting layers in OLEDs are PVD glass mixtures in which an emitter is dispersed in a host[17]. The molecular orientation of emitter molecules affects the emission of light from thin films and thereby the device efficiency. Recent work has demonstrated that a horizontal molecular orientation of the transition dipole of light emitters can increase device efficiency by at least a factor of 1.3, relative to random emitter orientations[18-20]. Previous work on emitter orientation has identified the shape of the emitter molecule[21, 22] and the glass transition temperature of the host[22-24] as key variables, and much of this work[23, 25, 26] is consistent with the surface equilibration mechanism. Another example of the importance of molecular orientation in multicomponent PVD glasses is the orientation of polar molecules which determines the surface charge (or giant surface potential, GSP[27]), and this can have a major influence on charge injection in devices[28, 29]. Recent work maximized the surface charge by adjusting substrate temperature and deposition rate, in qualitative accord with the surface equilibration mechanism[30, 31]. These examples illustrate that it is practically important to understand and control molecular orientation in two-component PVD glasses of organic semiconductors.

Here, we perform a thorough survey of the stability and molecular orientation of six binary vapor-deposited organic semiconductor glass mixtures with 50:50 mass concentration. This regime in the middle of the composition space is anticipated to be the most challenging to understand. Scheme 1 shows the chemical structures and calorimetric glass transition temperatures ($T_g$) of the studied organic semiconductors. To ensure the diversity of studied mixtures, the selected compounds cover a broad range of $T_g$ values (from 332K to 450K) and different molecular shapes including rod-, disk-, and sphere-shaped molecules. By using differential scanning calorimetry (DSC), we evaluate the enthalpy and kinetic stability of these co-deposited glass mixtures of organic semiconductors at substrate temperatures, $T_{sub}$=0.78-0.88$T_{g,mixture}$; this is the temperature window where the most stable single-component PVD organic semiconductor glasses are obtained. The results show that all six co-deposited glass mixtures exhibit high kinetic stability and low enthalpy, comparable to the most stable single-component PVD organic glasses. This high stability and low enthalpy is observed even when the difference in pure component $T_g$ values exceeds 90 K. In addition, the molecular orientation of the binary glasses deposited across a wide range of $T_{sub}$ is investigated using variable angle spectroscopic ellipsometry (VASE) to obtain the birefringence; in mixed glasses, the birefringence contains information about the orientation of both components. We find that, for all six binary mixtures, the birefringence is controlled by $T_{sub}/T_{g,mixture}$, akin to that in single-component PVD organic glasses. Furthermore, we show that the birefringence of a binary PVD glass mixture can be predicted from the birefringence of the neat PVD glasses of the two components, through a model derived from the surface equilibration mechanism.

**Scheme 1.** Chemical structures and glass transition temperatures ($T_g$) of the organic semiconductors studied here. The $T_g$ is determined from DSC measurement in a heating process with 10K/min (except for CBP, where 50K/min was utilized[32]). $T_g$ values for TPD and m-MTDATA is taken from ref.[16]. DSC results for DSA-Ph, TCTA, and Alq3 are given in **Supporting Information (SI)**.

## ▪ EXPERIMENTAL METHODS

**Materials.** TPD (N,N'-Bis(3-methylphenyl)-N,N'-diphenylbenzidine, 99%), CBP (4,4'-Bis(N-carbazolyl)-1,1'-biphenyl, 99.9%), m-MTDATA (4,4',4''-Tris[phenyl(m-tolyl)amino]triphenylamine, 98.7%), TCTA (Tris(4-carbazoyl-9-ylphenyl)amine, >97%), and $Alq_3$ (Tris-(8-hydroxyquinoline) aluminum, 99.9%) were purchased from Sigma-Aldrich. DSA-Ph (1-4-Di-[4-(N,N-diphenyl)amino]styryl-benzene, 99%) was purchased from Luminescence Technology Corporation. All compounds were used without further purification.

**PVD glass mixture preparation.** Co-deposited glass mixtures were prepared in a vacuum chamber with a base pressure ~$10^{-6}$ Torr. The deposition rate for each component was controlled individually by heating two independent crucibles using resistive wire heaters. By setting the same deposition rate for the two components, we prepared PVD glass mixtures with a 50:50 mass ratio of components. The total deposition rate was 0.42±0.02 nm/s and monitored using a quartz crystal microbalance (QCM). The thickness of the deposited films was measured by QCM. The substrate temperature was held constant during deposition using a Lakeshore controller with platinum RTD sensors. Co-deposited glassy films with a thickness of 1400 nm were deposited onto 120nm thick gold foil (purchased from Barnabas Gold) for differential scanning calorimetry measurements, while films with a thickness of 380-400nm for spectroscopic ellipsometry measurements were deposited onto one-side polished silicon wafers (purchased from Virginia Semiconductor).

**Differential scanning calorimetry (DSC) measurements.** Thermal analysis of bulk and co-deposited samples was performed using a TA Q2000 differential scanning calorimeter (New Castle, DE). To determine the kinetic stability and enthalpy of co-deposited glass mixtures, the as-deposited films with the attached gold foil (120nm thickness) were folded and loaded into a Tzero pan; the pan was sealed by a Tzero lid using a crimper press to achieve good contact between the tested sample and the pan. The scanning rate was 10K/min for both heating and cooling processes under 50mL/min $N_2$ purge.

**Spectroscopic ellipsometry measurements.** The optical properties of co-deposited glass mixtures were probed using spectroscopic ellipsometry (J.A. Woollam M-2000U). The optical parameters $\Psi$ and $\Delta$ were determined in the wavelength range of 600-1000 nm at room temperature through variable angle measurements at three incidence angles of 50°, 60°, and 70°. The anisotropic Cauchy model was applied to model the experimental data, determining the thickness and the ordinary ($n_0$) and extraordinary ($n_e$) refractive indices. The birefringence ($\Delta n = n_e - n_0$) of co-deposited glass mixtures is reported at a wavelength of 632.8 nm.

## ▪ RESULTS

**Kinetic stability.** We employ differential scanning calorimetry (DSC) to evaluate the kinetic stability of co-deposited glass mixtures of organic semiconductors. The obtained DSC results for six 50:50 binary systems are shown in Figure 1. The data are obtained during heating with a rate of 10 K/min. The pink and blue curves present the calorimetric data for the as-deposited glass mixtures. After the as-deposited glasses were entirely transformed into the supercooled liquid state, the samples were cooled at 10 K/min to form liquid-cooled (LC) glasses. The gray curves denote the reheating DSC results of the corresponding LC glass mixtures, from which the glass transition temperature of the mixtures, $T_{\mathrm{g,mixture}}$, was determined and displayed in Figure 1. Notably, the $T_{\mathrm{g,mixture}}$ values obtained here are in good agreement with those of their corresponding bulk mixtures with 50:50 mass ratio, as shown in Figure S2. The DSC result of bulk TPD/m-MTDATA mixture can be found in ref.[16]. For each panel in Figure 1, the heat capacity is normalized to the heat capacity change, $\Delta C_{\mathrm{p}}$, of the liquid-cooled glass mixtures during the glass to liquid transition.

For every binary system of organic semiconductors considered, the kinetic stability of the co-deposited glasses prepared at $T_{\mathrm{sub}}$=0.78-0.88$T_{\mathrm{g,mixture}}$ is significantly enhanced, in comparison to the corresponding liquid-cooled glass. As shown in Figure 1, for each system, the onset temperature $T_{\mathrm{onset}}$, where the PVD glass mixture starts to transform, is 15-21 K higher than the $T_{\mathrm{g,mixture}}$ of the liquid-cooled glass. This is a straightforward indication that the co-deposited glass mixtures prepared at these substrate temperatures are kinetically more stable since more thermal energy is required to dislodge the molecules from the solid-state packing formed by vapor deposition. For one of the six mixtures, a slightly different procedure was used; since the as-deposited DSA-Ph/Alq3 mixture crystallized during the glass-to-liquid transformation process, the $T_{\mathrm{g,mixture}}$ and $\Delta C_{\mathrm{p}}$ values of DSA-Ph/Alq3 mixture could not be obtained directly. We estimate $T_{g,mixture}$=392.5 K for DSA-Ph/Alq3 at a 10K/min heating rate (4K higher than the previously reported value obtained at 1K/min[33]) and $\Delta C_p$ = 0.369 K $J^{-1}$ $g^{-1}$ (the weighted average of $\Delta C_p$ for pure DSA-Ph and pure Alq3). The detailed procedure for estimating the heat capacity for the DSA-Ph/Alq3 liquid is given in **Supporting Information (SI)**. We infer that all as-deposited mixtures studied here form a single homogenous glassy phase and the presence of two peaks (or usually a shoulder on a single peak) is not due to phase segregation in components based on two reasons: 1) these peaks/shoulders are not completely reproducible. One example has been presented in Figure S4; 2) the DSC response spans a narrow temperature interval around 8-15K, which is comparable to (but slightly larger than) the glass transition width for single-component PVD glasses (Figure 2). The peaks/shoulders shown in the DSC data of certain systems arise from variations in composition within layers due to the fluctuations in individual deposition rates and the composition change across films due to the deposition chamber geometry, such that it may be 48:52 at one end and 52:48 at the other end.

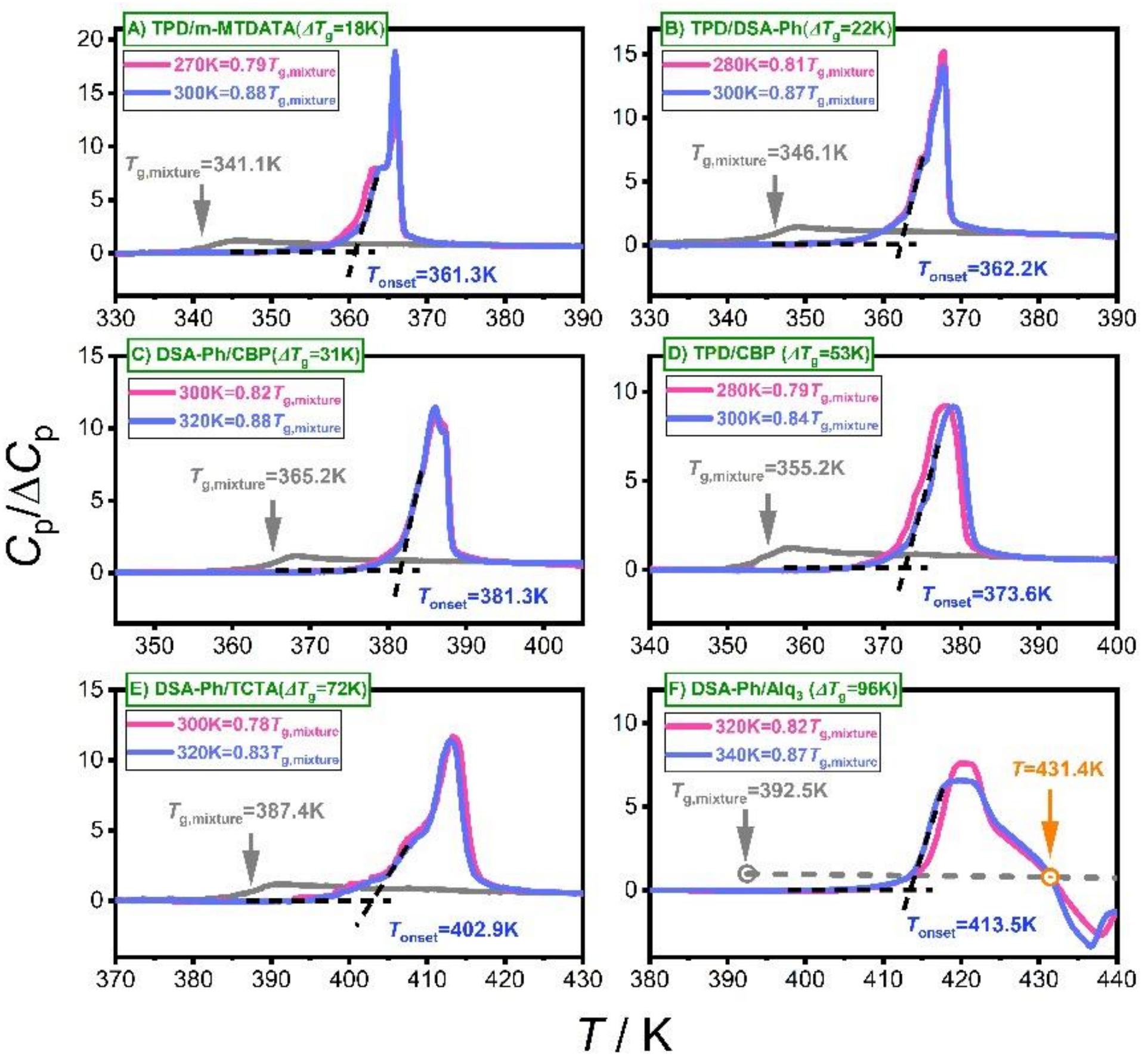

**Figure 1.** DSC results for six 50:50 binary mixtures of organic semiconductors. The pink and blue curves present the results of co-deposited glass mixtures with the substrate temperatures given in the legend, and gray curves denote results for the corresponding liquid-cooled glass mixtures. For panel F, the heat capacity of the supercooled liquid of DSA-Ph/Alq3 is predicted according to the procedure given in SI.

In Figure 2, we compare the kinetic stability of these co-deposited glass mixtures with the most stable glass of pure TPD[16] (deposited at $T_{sub}$=0.86$T_g$ and shown in grey); this TPD data is representative of that obtained for single component PVD glasses of organic semiconductors[16]. For this purpose, we normalize the scanning temperature to the glass transition temperature of the liquid-cooled glasses. Remarkably, for all six binary systems investigated, the temperature where the vapor-deposited glass mixtures start to transform is around 1.05 times higher than the corresponding $T_g$ value. As shown, this kinetic stability is comparable to that of the most stable vapor deposited TPD glass. It is worth noting that the difference the glass transition temperatures of the two components, $\Delta T_g$, is as large as 96K, which is five times larger than a recent work[16].

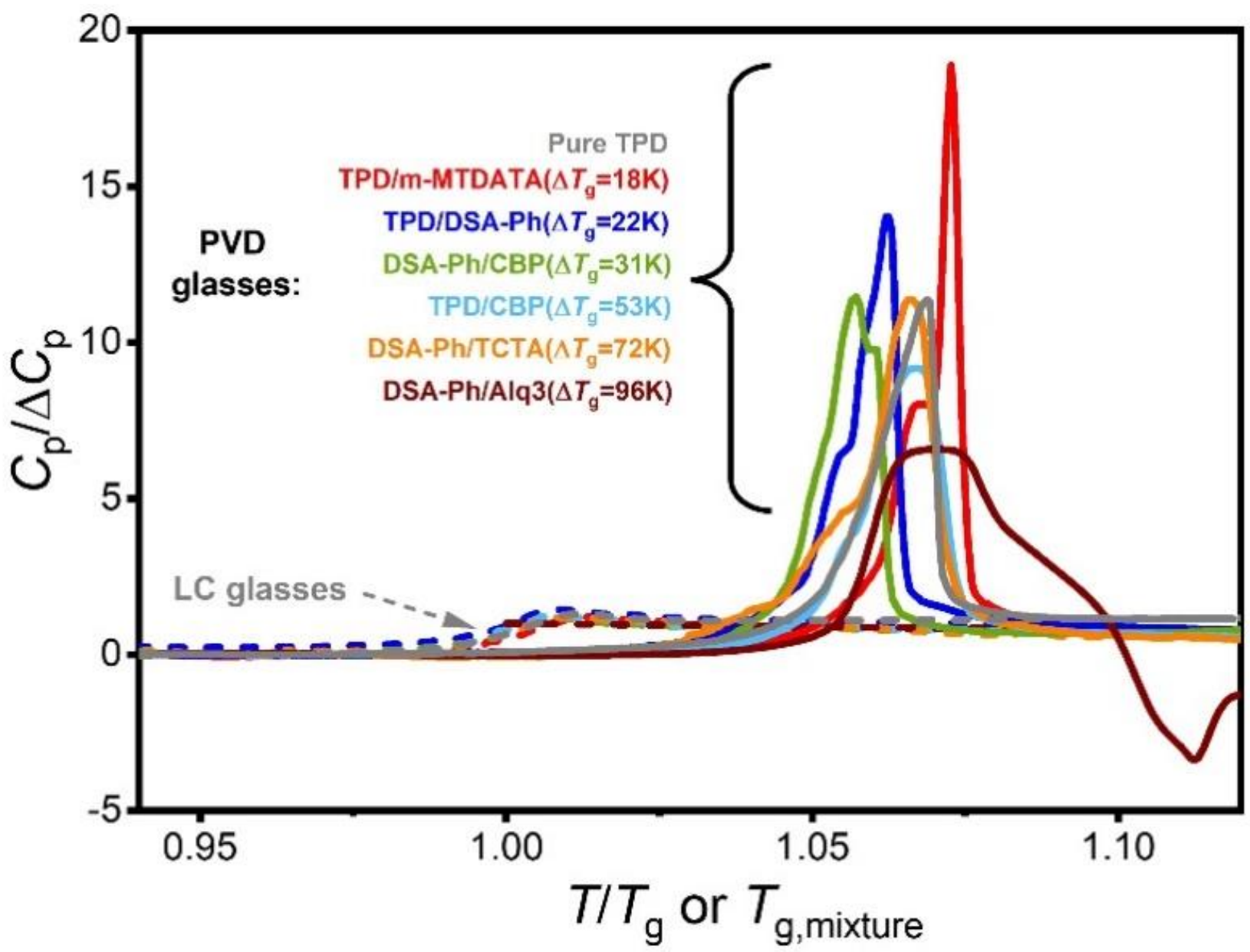

**Figure 2**. DSC heating results for vapor-deposited (solid lines) and liquid-cooled (dashed lines) organic semiconductor glass mixture. The scanning temperature is normalized to the glass transition temperature of the samples. Note, for each mixture, the DSC data of the sample deposited at the higher $T_{sub}$ is displayed.

**Enthalpy.** As shown in Figure 2, the co-deposited glasses show a pronounced endothermic peak during the glass-to-liquid transition, while their corresponding liquid-cooled glasses exhibit a step-like change in heat capacity with a negligible endothermic peak. Similar to single-component systems, this is an indication that the co-deposited PVD glass mixtures are low in energy landscape, akin to highly aged glasses[5]. The enthalpy of each PVD glass is obtained by integrating the heat capacity curves in Figure 1, and these results are shown in Figure 3. For all the studied systems, the enthalpy of co-deposited glass mixtures is significantly lower than that of the corresponding liquid-cooled glass mixtures.

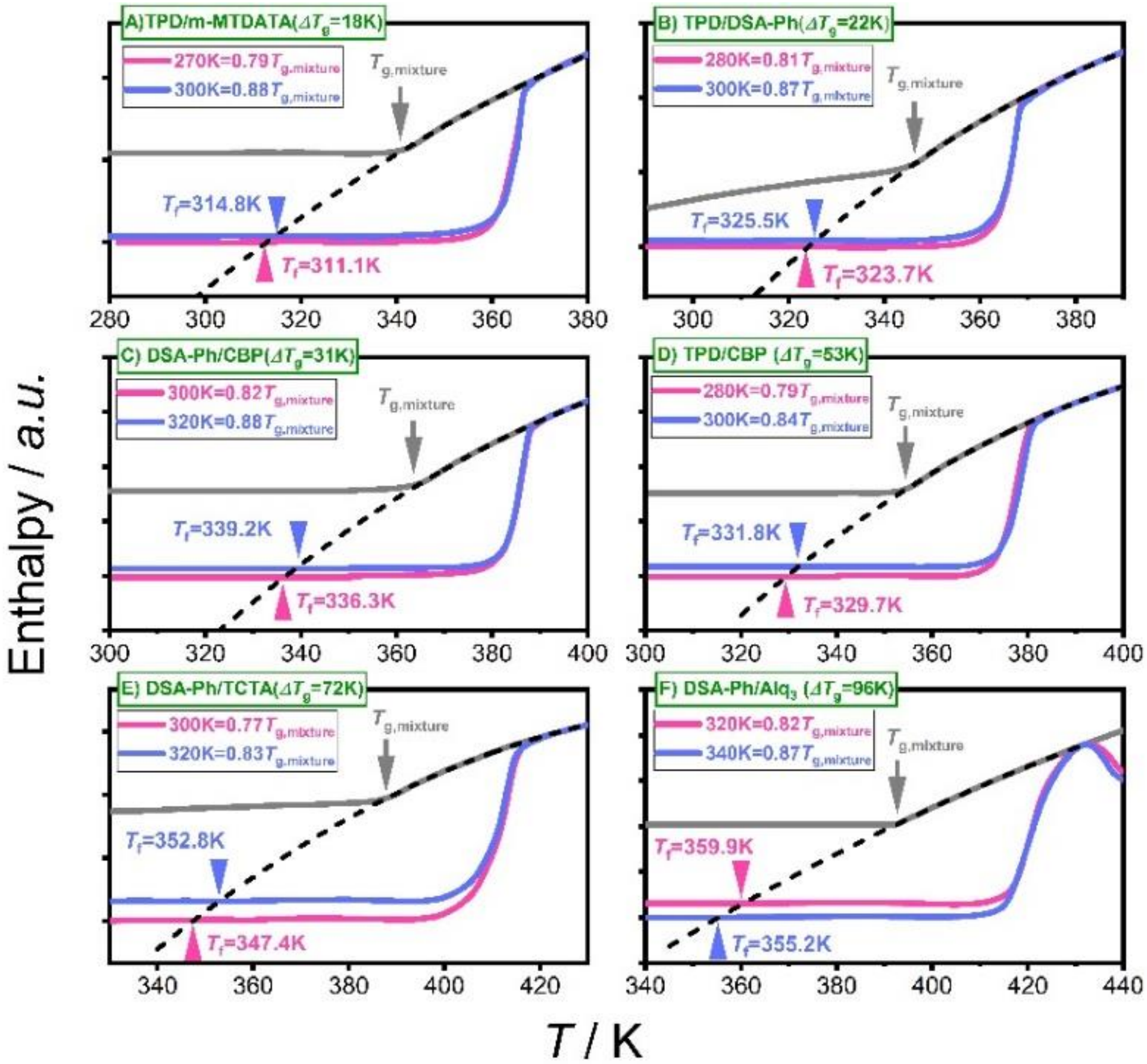

**Figure 3.** Enthalpy is plotted as a function of temperature for six co-deposited glass mixtures (pink and blue) and corresponding liquid-cooled glass mixtures (gray). Dashed lines represent quadratic fits to the enthalpy data in the equilibrium liquid and extrapolation to lower temperature. Fictive temperatures ($T_f$) are determined by the intersection of the PVD glass enthalpy with the extrapolated supercooled liquid enthalpy.

We compare the enthalpies of different PVD glasses through the use of the fictive temperature $T_f$, and we find that $T_f$ for each of the binary PVD glasses is much lower than $T_{g,mixture}$. The fictive temperature $T_f$ is frequently used to determine the extent to which a glass is equilibrated. We determine $T_f$ as the temperature where the enthalpy of the PVD glass meets the extrapolation of the supercooled liquid enthalpy (dashed lines in Figure 3). The $T_f$ values for the PVD glasses are around 22-40 K lower than $T_{g,mixture}$, indicating extraordinary thermodynamic stability in co-deposited mixtures. The uncertainty in the determined $T_f$ values is ±3K, resulting from uncertainty in extrapolating the supercooled liquid enthalpy to low temperatures. For context, we note that $T_f$ for the most stable single-component PVD glasses (*i.e.*, TPD[16, 34], TNB[5], and IMC[35]) is about 30 K lower than their $T_g$, and similar results have been reported for amber glasses aged for tens of millions of years[36, 37].

**Comparison with single-component PVD organic glasses.** The enhanced kinetic stability and reduced enthalpy of these co-deposited glass mixtures of organic semiconductors are comparable to the most stable single-component organic glasses. As shown in Figures 4A and 4B, for all co-deposited glass mixtures, the determined $T_{onset}/T_{g,mixture}$ values are between 1.04-1.06 (the gray shaded region in Figure 4A), and simultaneously the $T_f/T_{g,mixture}$ values are in range of 0.90-0.94 (the gray shaded region in Figure 4B). These values are very similar to those for vapor-deposited single-component systems (*e.g*. TPD shown here) prepared in the same $T_{sub}/T_g$ regime[13, 38]. This strongly supports the view that the kinetic stability and enthalpy of PVD glass mixtures and neat glasses are both controlled by the surface equilibration mechanism. Furthermore, our results show that ultrastable glass mixtures are generally obtained when deposited around $0.85T_{g,mix}$, regardless of the molecular shape and $T_g$ difference of the two components.

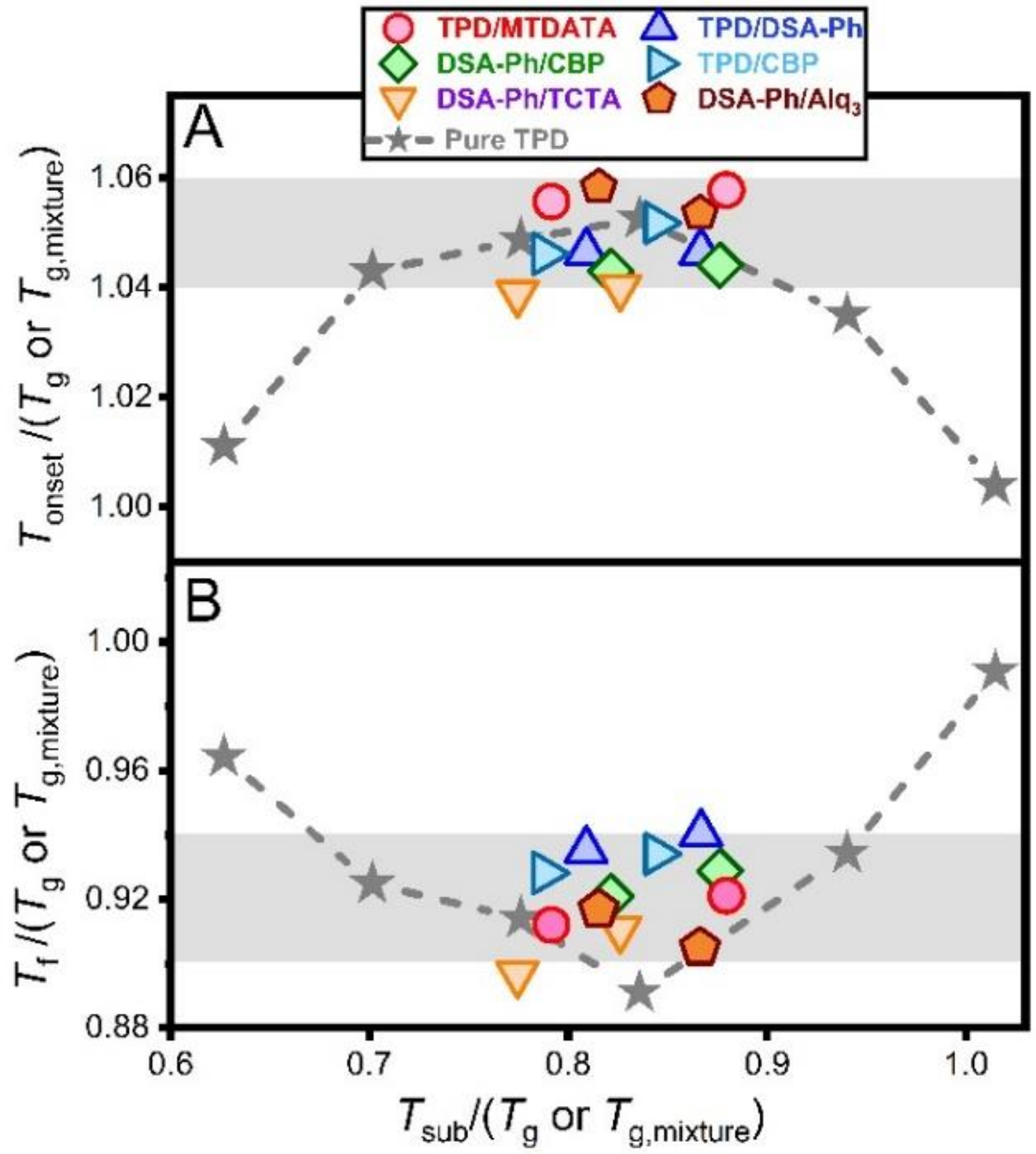

**Figure 4.** The kinetic stability (A) and enthalpy (B) of co-deposited glass mixtures of organic semiconductors, in comparison to a single-component system. Results are presented as a function of substrate temperature during deposition. The dashed lines connect data for single-component PVD glasses of TPD.

**Molecular Orientation.** We employed spectroscopic ellipsometry to study the average molecular orientation in the six co-deposited glass mixtures. The birefringence provides an effective way to evaluate the average molecular orientation and is defined as $\Delta n = n_e - n_0$ (where $n_e$and $n_0$ are the extraordinary and ordinary indices of refraction, respectively). Figure 5 shows the birefringence (red spherical points) determined from spectroscopic ellipsometry for these binary organic semiconductor glasses co-deposited at $T_{sub}/T_{g,mixture}$=0.75-1.02.

All the binary mixtures show similar trends in birefringence when plotted as a function of $T_{\mathrm{sub}}/T_{\mathrm{g,mixture}}$. At lower values of $T_{\mathrm{sub}}/T_{\mathrm{g,mixture}}$, all co-deposited mixtures exhibit significant negative birefringence, indicating a tendency towards "face-on" packing. In contrast, isotropic PVD glass mixtures with $\Delta n \approx 0$ are obtained when $T_{\mathrm{sub}}/T_{\mathrm{g,mixture}}$ is very near unity. This indicates that $T_{\mathrm{sub}}/T_{\mathrm{g,mixture}}$ is a key factor in controlling the molecular orientation of co-deposited binary organic semiconductor glasses. All binary mixtures studied here have a 50:50 mass ratio of the two components, except for the DSA-Ph/Alq3 mixtures with 58%DSA-Ph and 42%Alq3, as reported in ref.[33].

We find that the birefringence of the co-deposited glass mixtures can be quantitatively predicted. Specifically, for a given glass mixture AB, the observed birefringence is the weighted average of the pure glasses' birefringence obtained under the "iso-mobility" deposition condition $\frac{T_{sub,AB}}{T_{g,AB}} = \frac{T_{sub,A}}{T_{g,A}} = \frac{T_{sub,B}}{T_{g,B}}$:

$$\Delta n_{AB}\left(\frac{T_{sub,AB}}{T_{g,AB}}\right) = \varphi_A * \Delta n_A\left(\frac{T_{sub,A}}{T_{g,A}}\right) + \varphi_B * \Delta n_B\left(\frac{T_{sub,B}}{T_{g,B}}\right) \tag{1}$$

where $\Delta n_{AB}$, $\Delta n_A$, and $\Delta n_B$ denote the birefringence of PVD glass mixture AB, pure glass A, and pure glass B; and $\varphi_A$ and $\varphi_B$ are the volume fractions of component A and component B in the mixture. Eq. 1 does not simply compute the average birefringence for a given substrate temperature (see Discussion below). Rather, as an example, the birefringence of a mixture deposited at $0.8T_{\mathrm{g,mixture}}$ is computed from the birefringence of component A deposited at $0.8T_{\mathrm{g,A}}$ and the birefringence of component B deposited at $0.8T_{\mathrm{g,B}}$. In our analysis, we assume that the volume fraction is equal to the weight fraction for PVD organic glasses.

Remarkably, the calculated birefringence (black solid lines in Figure 5) matches quite well with the experimental determinations (red spheres) for all six co-deposited organic semiconductor glass mixtures. This result again emphasizes that $T_{sub}/T_{g,mixture}$ is a key factor controlling the molecular orientation of co-deposited organic semiconductor glass mixtures, in analogy to single-component PVD organic glasses.

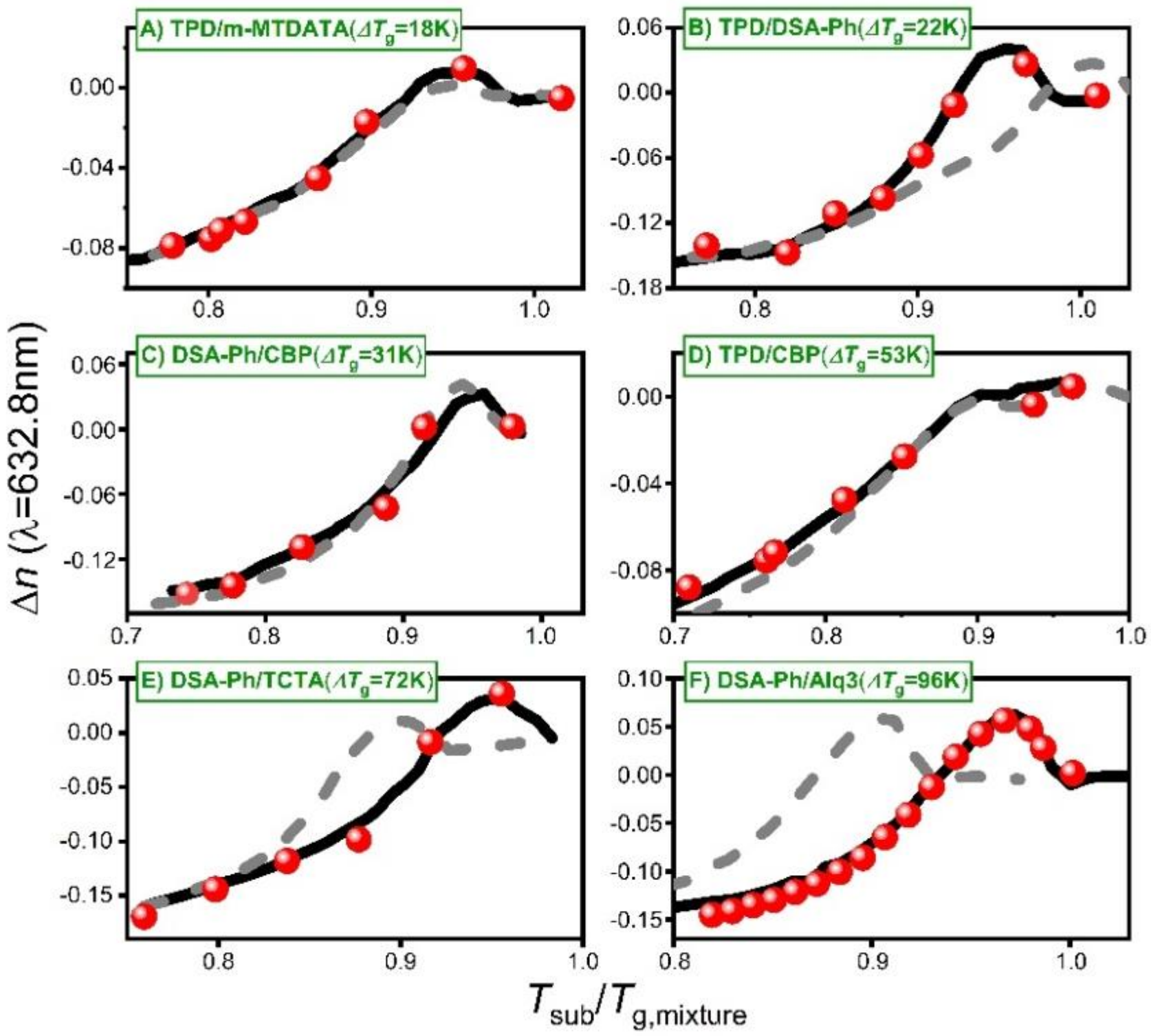

**Figure 5.** Birefringence of co-deposited glass mixtures of organic semiconductors at wavelength λ=632.8 nm. The red spheres denote the experimental birefringence determined from spectroscopic ellipsometry, while the black solid lines represent the predicted birefringence from Eq. 1. The gray dashed lines represent the predicted birefringence from Eq. 2. The birefringence data of single-component PVD glasses used to perform these calculations are displayed in Figure S4.

## Discussion

**Stability of co-deposited semiconductor glasses.** Here we studied co-deposited organic semiconductor mixtures with various molecular shape combinations, and with $\Delta T_g$ for the two components up to 96K. We found that all six glass mixtures exhibit highly enhanced kinetic stability and significantly reduced enthalpy when co-deposited at $T_{sub}$=0.78-0.88$T_{g,mixture}$, comparable to the most stable single-component organic glasses. This is the first experimental evidence that stable glass formation is general for organic semiconductor mixtures, even when $\Delta T_g$ is nearly 100 K.

These new results can be successfully interpreted by extending the surface equilibration mechanism[5] to account for surface mobility in mixed systems. The surface equilibration mechanism emphasizes that high molecular mobility within 1-2 nm of the surface is the key for forming ultrastable PVD glasses, since this allows molecules to reach (or nearly reach) equilibrium states at temperatures lower than the glass transition temperature before they are buried by further deposition. For a co-deposited mixture, we expect that *both* components must have high surface mobility in order to form an ultrastable glass. In a previous publication[16], we assumed that the surface mobility of a component is only determined by temperature (and not influenced by composition). We showed that this scenario could (just barely) explain ultrastable glass formation for $\Delta T_g$ = 18 K. However, this scenario cannot explain ultrastable glass formation when $\Delta T_g$ is large, since the high-$T_g$ component would have very little mobility at 0.85$T_{g,mixture}$. To explain our results, we infer that the surface mobility of each component is strongly influenced by composition, in addition to temperature, so that the two components have similar surface relaxation times (*i.e.*, similar mobility) on top of the co-deposited glass. By way of analogy, we note that the bulk glass transition has this character. If two molecules with different $T_g$ values form a single liquid phase, a single, intermediate $T_g$ value is usually observed[39], indicating that the two molecules have similar mobilities in the mixture. With this inference that the surface mobility of each component is similar on top of the co-deposited glass, the kinetic and thermodynamic stability of well-mixed co-deposited glasses are naturally very similar to the behavior of ultrastable, single-component PVD glasses, as shown in Figure 4. The surface mobility of mixed glasses has rarely been studied[40], and an experimental test of our inference is an important goal for future work.

The results of the current study complement and extend the five previous literature reports of the kinetic and thermodynamic stability of multicomponent organic glasses. Two of these studies showed that mixtures of isomers (*cis*-/*trans*-decahydronaphthalene[41] and cis-/trans-decahydroisoquinoline[15]) with identical glass transition temperatures can form highly stable glasses. Qiu *et al.*[42] showed that vapor-deposited dilute glass mixtures of 5% 4,4-diphenylazobenzene/95% celecoxib can show highly increased density and enhanced kinetic stability as compared to the liquid-cooled glass mixture. Our present work generalizes these results to mixtures with large $\Delta T_g$ and outside the dilute regime. Recently, two works studied the stability of PVD glass mixtures of organic semiconductors. Cheng *et al.*[16] reported that 50:50 co-deposited glass mixtures of TPD and m-MTDATA formed highly stable glasses. The authors pointed out that small values of $\Delta T_g$ (=18 K) and the ideal solution behavior between these two organic semiconductors were likely reasons why this binary system behaves like a neat PVD glass; our present work shows that a small $\Delta T_g$ is not a prerequisite. Ki *et al.*[43] reported that co-deposited glass mixtures of 8-hydroxyquinolinolato-lithium (Liq) and 4,7-Diphenyl-1,10-phenanthroline (BPhen) prepared at $T_{sub}$=0.80-0.89$T_{g,mixture}$ did not show ultrastable behavior. We note that they also reported that pure Liq failed to form stable glasses via PVD. Inability of Liq to form stable glasses via PVD could be interpreted as a lack of surface mobility, which then might explain why

Liq/BPhen mixtures do not form stable glasses. This is a key difference from the six mixtures studied here whose components can individually form stable glasses via PVD.

**Molecular orientation in co-deposited semiconductor glasses.** In Figure 5, we demonstrate that $T_{sub}/T_{g,mixture}$ is the key factor controlling the birefringence of PVD organic semiconductor glass mixtures. We studied six co-deposited systems in a broad substrate temperature range from $0.75T_{g,mixture}$ to $1.02T_{g,mixture}$, in which molecular packing varies continuously from "face-on" to "edge-on" and to isotropic packing.

The observation that Eq. 1 can successfully predict the birefringence of the co-deposited mixtures is significant and we want to specify the two assumptions needed to derive this equation. First, we assume that mobility of a given molecule at the surface of the co-deposited glass is determined only by $T_{sub}/T_{g,mixture}$, for any composition of the glass. This assumption naturally leads to the conclusion that the two components will have similar mobilities during co-deposition, as we inferred above based upon the ultrastability of co-deposited mixtures with large $\Delta T_g$. Second, we assume that a given surface mobility always leads to the same molecular orientation at the surface, independent of the composition of the surface. We have no independent argument for the validity of this assumption, beyond the observation that Eq. 1 is quite accurate, and we do not know how to derive this equation without this assumption. Eq. 1 is a generalization of a relation proposed earlier by Jiang *et al.*[33] to explain the birefringence of the PVD glasses of DSA-Ph/Alq3 where PVD glasses of pure Alq3 have no birefringence[44].

We wish to emphasize that Eq.1 is not a simple statement that the co-deposited mixture at a specific $T_{sub}$ will have the average birefringence of the two pure components deposited at the same $T_{sub}$. We show here that such a simple average does not describe the experimental data. If a simple average were valid, then at a given $T_{sub}$, we would have

$$\Delta n_{AB}(T_{sub}) = \varphi_A * \Delta n_A(T_{sub}) + \varphi_B * \Delta n_B(T_{sub}) \tag{2}$$

A comparison of our experimental data (the red spheres) with the birefringence predicted from Eq. 2 (the gray dashed lines) is shown in Figure 5. While some systems are reasonably described by Eq. 2, at least three mixtures are poorly described. The success of Eq. 1 and the failure of Eq. 2 indicates the key role for $T_{sub}/T_{g,mixture}$ in determining molecular orientation in co-deposited glasses, which in turn signals the importance of mobility for this process, as anticipated by the surface equilibration mechanism. In more detail, it suggests that the two assumptions used to derive Eq. 1 are correct, at least for organic semiconductor mixtures similar to the six systems studied here. The success of Eq. 1 also implies that the birefringence contribution of each component in the mixture is individually described by our approach. Thus, we can also use this approach to understand the molecular orientation of each component in the mixture.

Our work complements and extends recent studies on the molecular orientation of emitters in PVD glass mixtures of organic semiconductors, with the goal of increasing device efficiency. Brutting *et al*. studied the orientation of a coumarin dye[23] and four nonpolar dyes[45] in a series of dye-doped guest-host systems deposited at room temperature and observed increasing horizontal orientation of the dye with decreasing $T_{sub}/T_{g,host}$. Since $T_{g,host}$ is expected to be quite close to $T_{g,mixture}$ in case of dilute mixtures, their work indicates that the $T_{sub}/T_{g,mixture}$ controls the molecular orientation of emitters in dilute mixtures. Moreover, utilizing three different substrate temperatures, Komino *et al.*[25] observed that a linear dopant molecule in a host matrix of 3,3'-di(9H-carbazol-9-yl)-1,1'-biphenyl (mCBP) had a greater tendency to orient horizontally at lower substrate temperatures. Although these previous papers focused on dilute mixtures and studied at most three substrate temperatures, their results are consistent with our conclusion that $T_{sub}/T_{g,mixture}$ is the key factor controlling molecular orientation in vapor-deposited glass mixtures. Because a wide range of substrate temperatures and non-dilute mixtures were investigated in our work, we uncovered a larger pattern in molecular orientation that extends to both components and applies to the entire range of possible compositions.

The result reported here may be useful for understanding the surface potential of vapor-deposited organic semiconductors, which plays a key role in controlling charge injections[46-48]. When polar molecules are deposited, any net alignment of the dipole moments (relative to the surface normal) can lead to a giant surface potential (GSP), which can exceed 15 V for 100 nm of deposited glass[49]. Recent work by Adachi and coworkers[50] showed that fluorine groups can be used to direct dipole orientation at the surface during deposition, leading to the production of glass films with either positively or negatively charged surfaces. As indicated in ref.[50], this result strongly supports the surface equilibration mechanism, at a qualitative level. Very recently, He *et al.*[30] reported that the deposition rate can be used control the GSP of pure organic semiconductors and 50:50 mixtures, in qualitative accord with the surface equilibration mechanism. These authors reported that the GSP for co-deposited glasses was approximately the average GSP for pure materials. We speculate that, over a broad range of deposition conditions, a more accurate prediction for GSP for mixtures might be obtained by averaging GSP values at the same value of $T_{sub}/T_{g,mixture}$, in analogy to Eq. 1. In some studies, mixtures of polar and non-polar molecules have been used to optimize GSP[51], and thus our observation that polar/non-polar pairs of molecules form ultrastable glasses is relevant for future work. It should be noted that GSP and birefringence characterize different aspects of anisotropic packing, even in the case where the dipole moment is aligned with the polarizability tensor. In our view, the surface equilibration mechanism should generally describe all measures of anisotropy in PVD glasses.

**Generality of these results.** Given the importance of co-deposited PVD glasses of organic semiconductors, it is useful to consider how general our results may be. We emphasize that the mixtures of organic semiconductors studied here are diverse: 1) covering various combinations of molecular shapes, *e.g.*, TPD/m-MTDATA (rod/disk), TPD/DSA-Ph (rod/rod), and DSA-Ph/Alq3 (rod/sphere); 2) spanning a wide range of $\Delta T_g$ in components from 18K (TPD/m-MTDATA) to 96K (DSA-Ph/Alq3); 3) including mixtures of non-polar/non-polar (TPD/CBP) and non-polar/polar (DSA-Ph/Alq3) molecules.

Based upon these results, we anticipate that if two organic semiconductors meet two key criteria, the co-deposited glass mixtures formed by them will exhibit ultrastability when deposited at $T_{sub}$=0.78-0.88$T_{g,mixture}$: 1) the organic semiconductors can form ultrastable glasses as pure components; and 2) the two components are well mixed in the glass (but without strong association). In addition, the molecular orientation of each component in the PVD glass mixtures can be predicted by Eq. 1. We anticipate that Eq. 1 is applicable to binary mixtures with compositions other than 50:50 (e.g., the dilute mixtures which are commonly utilized in OLEDs) if the two components mix well and show the ability to individually form ultrastable PVD glasses.

## ▪ Summary

The current work conducts a thorough investigation on the kinetic stability, enthalpy, and molecular orientation of six co-deposited 50:50 glass mixtures of organic semiconductors. The organic semiconductor mixtures studied here are diverse with $\Delta T_g$ ranging from 18K to 96K and covering various combinations of molecular shape and polarity. Nevertheless, all six systems exhibit high kinetic stability and significantly reduced enthalpy when deposited at $T_{sub}$=0.78-0.88$T_{g,mixture}$, compared to their corresponding liquid-cooled glassy mixtures. Furthermore, the birefringence of co-deposited organic semiconductor glass mixtures is quantitatively predictable based on the birefringence of the corresponding single-component PVD glasses using a mixing rule (Eq. 1) derived from the surface-equilibration mechanism.

We expect that these findings will extend to other co-deposited organic semiconductor glass mixtures, even those with compositions other than 50:50 (including dilute mixtures), if the two components can individually form PVD stable glasses and mix well during deposition. Consequently, a generic approach is proposed for producing highly stable PVD organic semiconductor glass mixtures and manipulating their molecular packing.

Physical vapor deposited glass mixtures of organic semiconductors are generally utilized as light-emitting layers in OLEDs. Our findings are significant for designing electronic devices with increased device lifetime and operational efficiency. Our work indicates that for a given organic semiconductor blend with room temperature deposition (*i.e.*, 298K), the most stable glass mixtures will be produced when the glass transition temperature of the mixture is around 340-370K, which is a criterion to manipulate the composition. Furthermore, our work shows that dilute mixtures with ultrastability for room temperature deposition will be obtained when the $T_g$ of host compounds is 340-370K. Previous work has shown that OLED devices prepared from ultrastable glasses can have substantially longer device lifetimes* and thus our new results provide guidance for device design. In addition, for a given pair of organic semiconductors, our work provides guidance on selecting an appropriate composition to obtain horizontal molecular orientation for room temperature deposition, which maximizes device efficiency.

## ▪ ASSOCIATED CONTENT

### Supporting Information

The following information is included: 1) DSC results for bulk organic semiconductors (Figure S1); 2) DSC results for 50:50 bulk mixtures of organic semiconductors (Figure S2); 3) Prediction of the specific heat capacity of 50:50 DSA-Ph/Alq3 mixture close to the glass transition (Figure S3); 4) Comparison of the DSC data of co-deposited TPD/DSA-Ph at $T_{sub}$=300K in two separate depositions; 5) Comparison of experimental $T_{g,mixture}$ of TPD/CBP and TPD/m-MTDATA with theoretical model; 6) The birefringence of single-component PVD glasses of studied organic semiconductors.

## ▪ AUTHOR INFORMATION

### Corresponding Author

*E-mail: ediger@chem.wisc.edu

### Author Contributions

The manuscript was written through contributions of all authors. All authors have given approval to the final version of the manuscript.

### Notes

The authors declare no competing financial interest.

## ▪ ACKNOWLEDGMENTS

Work by Shinian Cheng, Junguang Yu, Lian Yu, and Mark Ediger was supported by NSF through the University of Wisconsin Materials Research Science and Engineering Center (DMR-2309000). Work by Yejung Lee was supported by the U.S. Department of Energy, Office of Basic Energy Sciences, Division of Materials Sciences and Engineering, DE-SC0002161.

## ▪ REFERENCES

## Supporting Information for

# Generic behavior of ultrastability and anisotropic molecular packing in co-deposited organic semiconductor glass mixtures

*Shinian Cheng*[1], *Yejung Lee*[1], *Junguang Yu*[2], *Lian Yu*[2], *and M. D. Ediger**[,1]

[1]Department of Chemistry, University of Wisconsin-Madison, Madison, Wisconsin, 53706, USA

[2]School of Pharmacy, University of Wisconsin-Madison, Madison, Wisconsin, 53705, USA

**Corresponding Author**

*Email: ediger@chem.wisc.edu

This supporting information file includes:

1. DSC results for bulk organic semiconductors
2. DSC results for 50:50 bulk mixtures of organic semiconductors
3. Prediction of the specific heat capacity of 50:50 DSA-Ph/Alq3 mixture close to the glass transition
4. Comparison of the DSC data of co-deposited TPD/DSA-Ph at $T_{sub}$=300K in two separate depositions
5. Comparison of experimental $T_{g,mixture}$ of TPD/CBP and TPD/m-MTDATA with theoretical model
6. The birefringence of single-component PVD glasses of studied organic semiconductors

**1. DSC results for bulk organic semiconductors**

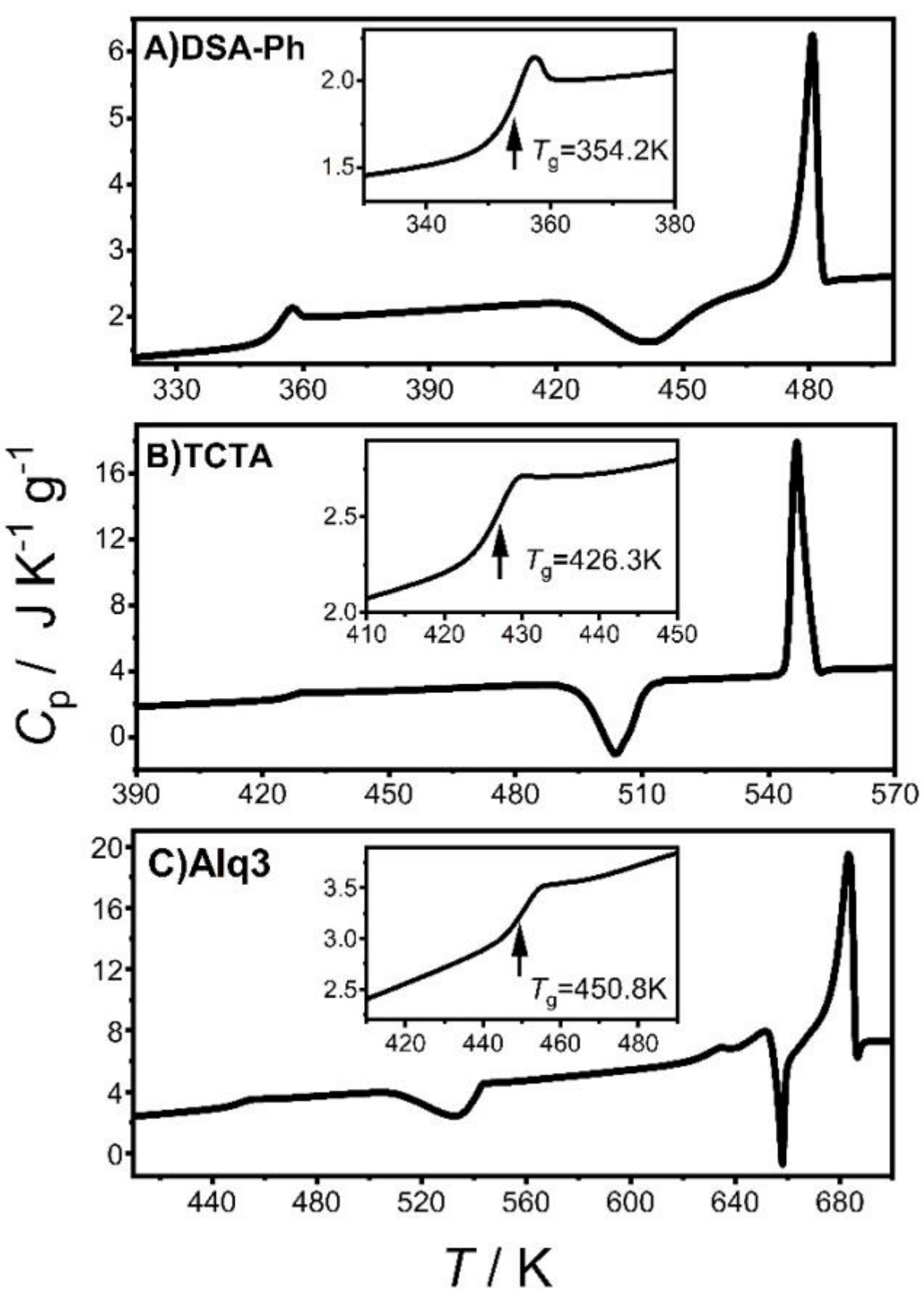

**Figure S1**. DSC results for bulk organic semiconductors: A) DSA-Ph, B) TCTA, and C) Alq3 measured in 10K/min heating process. The insets show the glass transition region.**2. DSC results for 50:50 bulk mixtures of organic semiconductors**

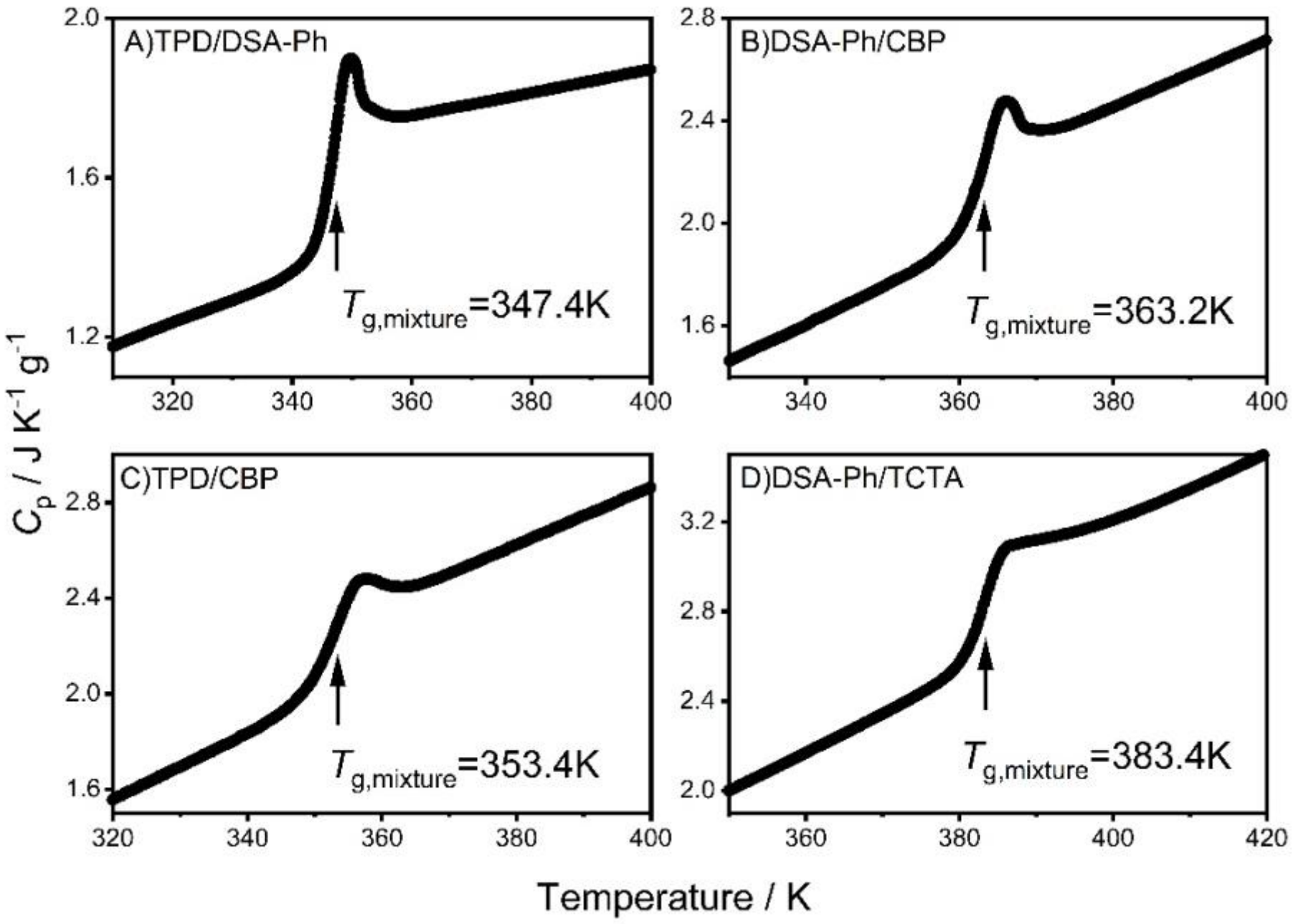

**Figure S2**. DSC results for 50:50 bulk mixtures of TPD/DSA-Ph (A), DSA-Ph/CBP (B), TPD/CBP(C), and DSA-Ph/TCTA (D) measured in 10K/min heating process. Since the melting point of Alq3 is 691K, which is 30K higher than the degradation temperature of DSA-Ph, we cannot prepare the bulk liquid mixture of DSA-Ph/Alq3 and thus we cannot obtain the corresponding liquid-cooled glass through DSC.

**3. Prediction of the specific heat capacity of 50:50 DSA-Ph/Alq3 mixture close to the glass transition**

To predict the specific heat capacity of 50:50 DSA-Ph/Alq3 mixture, $c_{p,mix}$, we assumed a linear dependence of $c_{p,mix}$ on temperature close to the glass transition. Under this assumption, if two heat capacity values are known for 50:50 DSA-Ph/Alq3 mixture at two specific temperatures, we can predict the heat capacity values in the entire temperature range.

In our analysis, the first heat capacity data point was selected at $T$=431.4K (orange circle in Figure 1F) where the DSC heating curves for the two as-deposited DSA-Ph/Alq3 mixtures intersect. It is worth noting that the two as-deposited mixtures should have same heat capacities in the supercooled liquid state since they have identical composition. The mass of the as-deposited DSA-Ph/Alq3 mixture is 0.60mg. The second heat capacity data point was selected at the glass transition temperature $T_{g,mixture}$=392.5K (gray circle in Figure 1F). To determine its value, we assumed that for a given 50:50 binary mixtures of organic semiconductors, the following relationship holds approximately:

$$\Delta c_{p,\mathrm{mix}} = \frac{1}{2}\left(\Delta c_{p,\mathrm{A}} + \Delta c_{p,\mathrm{B}}\right) \quad \text{(S1)}$$

where $\Delta c_{p,\mathrm{mix}}$, $\Delta c_{p,\mathrm{A}}$, and $\Delta c_{p,\mathrm{B}}$ are the specific heat capacity change of 50:50 binary mixture, pure component A, and pure component B during glass to liquid transition, respectively. Since $\Delta c_{p,\mathrm{DSA-Ph}}$ for pure DSA-Ph and $\Delta c_{p,\mathrm{Alq3}}$ for pure Alq3 can be well determined from DSC measurements as shown in Figure S3, the specific heat capacity change of 50:50 DSA-Ph/Alq3 mixture is calculated to be 0.369 K $J^{-1}$ $g^{-1}$. Based on these two data points, we generated the specific heat capacity data of DSA-Ph/Alq3 mixture by assuming a linear temperature dependence, as shown by the solid line in Figure 1F. To assess the reliability of these predictions, we applied the identical methodology to calculate the change in specific heat capacity for two additional mixtures and subsequently compared these results to the experimental data. For these two cases, the difference between the predicted and experimental values is less than 10%. If the error is similar for the DSA-Ph/Alq3 mixture, the resulting error in $T_f$ would be 2% (7K).

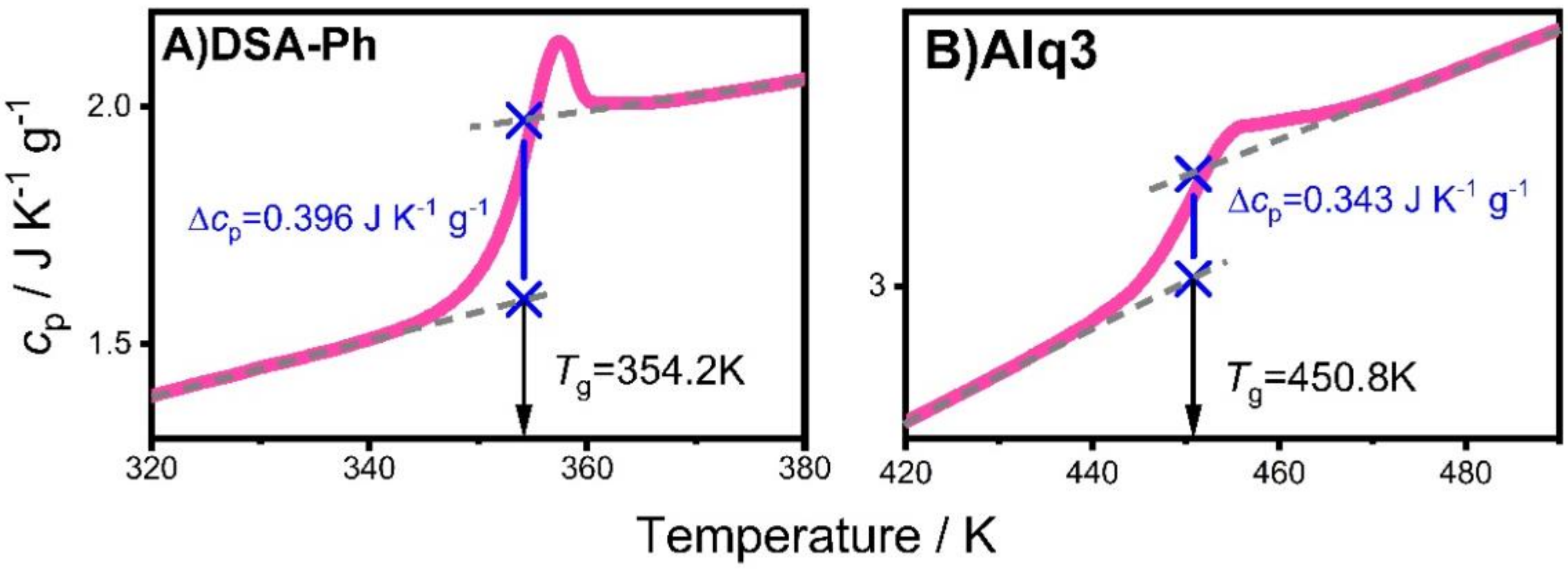

**Figure S3.** The DSC results for pure DSA-Ph (A) and pure Alq3 (B). The gray dashed lines are linear extrapolation of heat capacity data.

**4. Comparison of the DSC data of co-deposited TPD/DSA-Ph at $T_{sub}$=300K in two separate depositions**

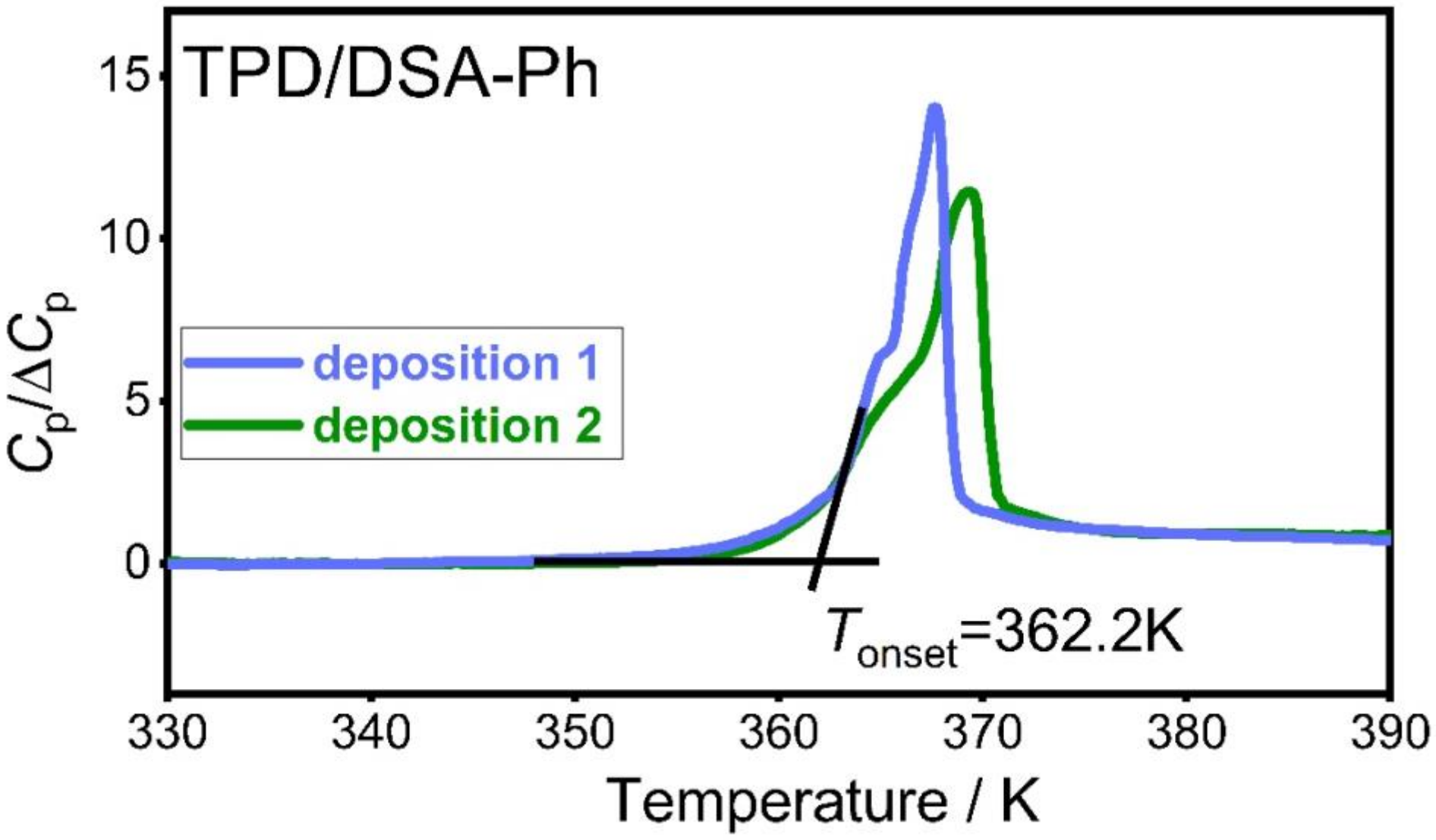

Figure S4. DSC result comparison of co-deposited TPD/DSA-Ph at $T_{sub}$=300K in two separate depositions. The $T_{onset}$ temperatures in two separate depositions are almost the same, but the peaks/shoulders in the glass transition range are not reproducible.

**5. Comparison of experimental $T_{g,mixture}$ of TPD/CBP and TPD/m-MTDATA with theoretical model**

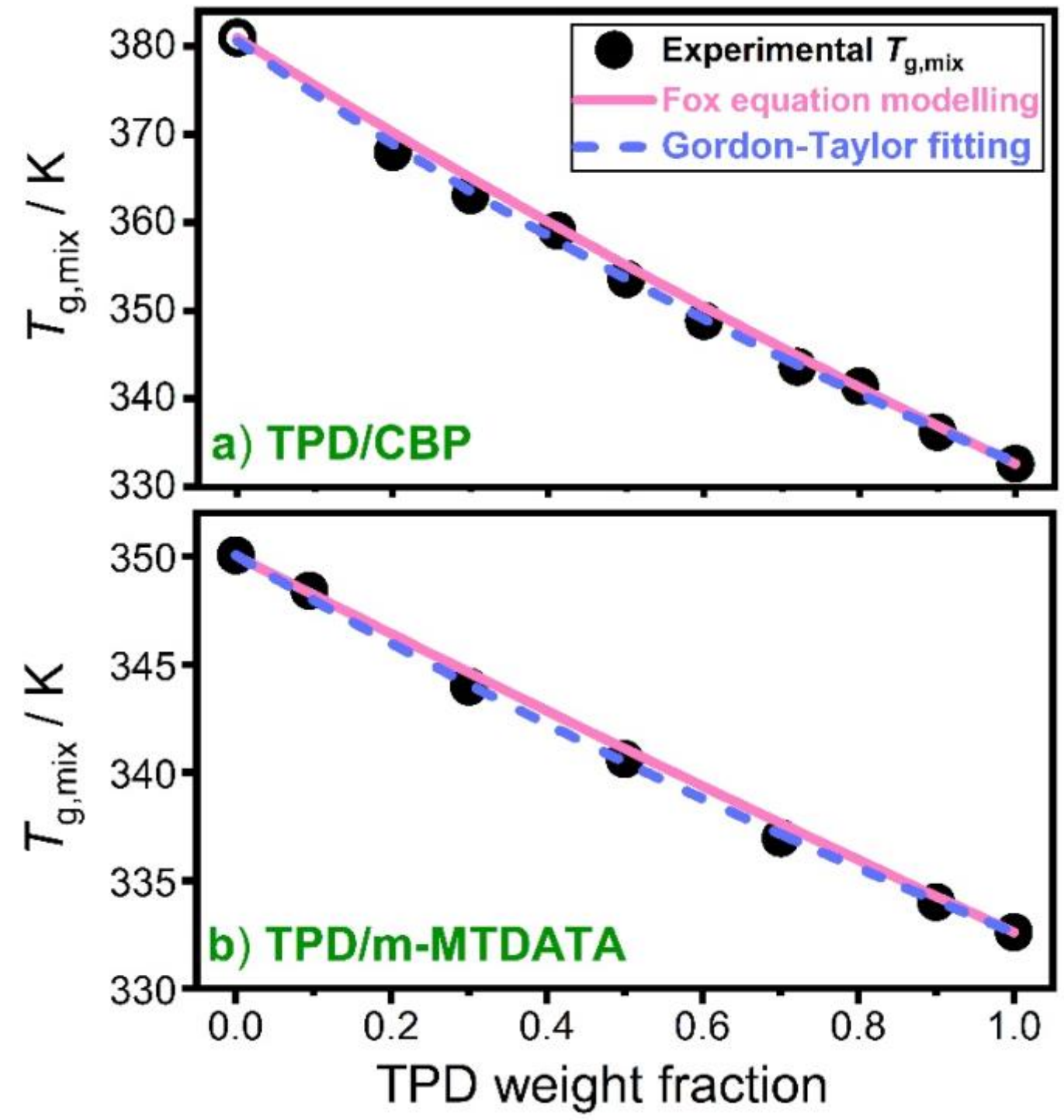

Figure S5. The comparison of the experimental glass transition temperature of (a)TPD/CBP and (b) TPD/m-MTDATA with the Fox equation prediction (solid pink line). In addition, the Gordon-Taylor equation fitting results (dashed blue line) are also presented as a comparison. The experimental Tg,mixture data is obtained from DSC with 10K/min heating rate. The glass transition temperature for 100% CBP at 10K/min is 381K (the open circle in graph a), estimated from its value ($T_g$=385K) determined at 50K/min heating rate[1].

## 6. The birefringence of single-component PVD glasses of studied organic semiconductors

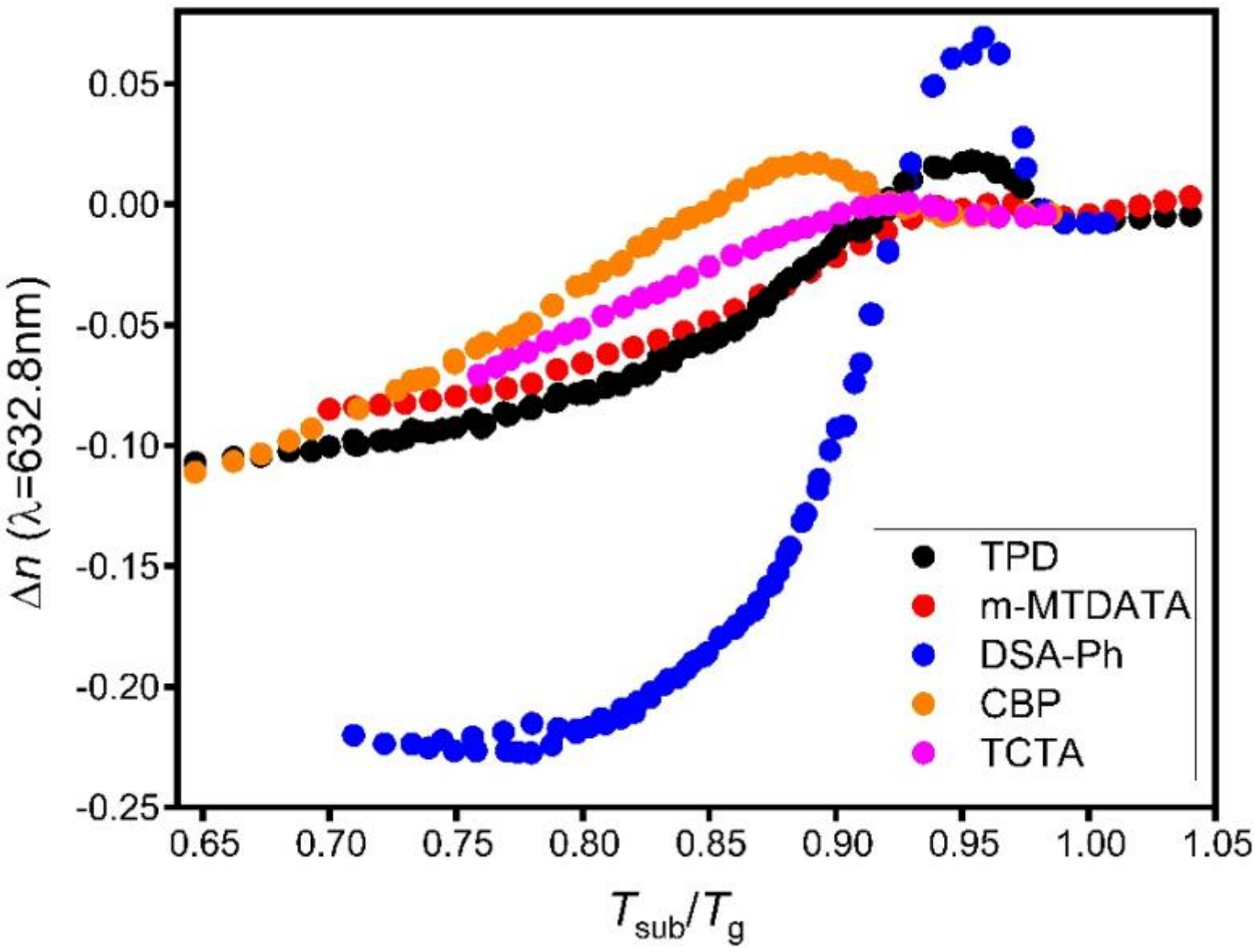

Figure S6. The birefringence of single-component PVD glasses of studied organic semiconductors with DSA-Ph and TPD taken from ref. [2], m-MTDATA and TCTA taken from ref. [3], and CBP from ref. [1]. The birefringence of vapor-deposited Alq3 is 0[4].